\documentclass[aps,prb,twocolumn,superscriptaddress,showpacs]{revtex4}
\usepackage{graphicx}

\usepackage{amsmath}
\usepackage{booktabs}
\usepackage{color}

\begin{document}
\title{On the longitudinal spin current induced by  a temperature gradient in a ferromagnetic insulator}
\author{S. R. Etesami}
\affiliation{Max-Planck-Institut f\"{u}r Mikrostrukturphysik, 06120 Halle, Germany}
\affiliation{Institut f\"{u}r Physik, Martin-Luther-Universit\"{a}t Halle-Wittenberg, 06120 Halle, Germany}
\author{L. Chotorlishvili}
\affiliation{Institut f\"{u}r Physik, Martin-Luther-Universit\"{a}t Halle-Wittenberg, 06120 Halle, Germany}
\author{A. Sukhov}
\affiliation{Institut f\"{u}r Physik, Martin-Luther-Universit\"{a}t Halle-Wittenberg, 06120 Halle, Germany}
\author{J. Berakdar}
\affiliation{Institut f\"{u}r Physik, Martin-Luther-Universit\"{a}t Halle-Wittenberg, 06120 Halle, Germany}

\begin{abstract}
Based on the solution of the stochastic Landau-Lifshitz-Gilbert equation discretized for a ferromagnetic chain subject to  a uniform temperature gradient, we present a detailed numerical study  of the spin dynamics with a focus  particularly  on finite-size  effects.
We calculate and analyze
the net longitudinal  spin current for various temperature gradients, chain lengths, and external static magnetic fields. In addition, we model an interface formed by a nonuniformly magnetized finite-size ferromagnetic insulator  and a normal metal and  inspect the effects of enhanced Gilbert damping on the formation of the space-dependent spin current within the chain. A particular aim of this study is the inspection of the spin Seebeck effect beyond the linear response regime. We find  that within our model the microscopic mechanism  of the spin Seebeck current is the magnon accumulation effect quantified in terms of the exchange spin torque. According to our results, this effect drives the spin Seebeck current  even in the absence of  a deviation between the magnon and phonon temperature profiles. Our theoretical findings are in line with the recently observed experimental results by M. Agrawal\emph{ et al.},  Phys. Rev. Lett. \textbf{111}, 107204 (2013).
\end{abstract}
\date{\today}

\maketitle

\section{Introduction}

Thermal magneto- and electric effects have a long history and are  the basis for a wide range of contemporary  devices.
%
Research activities revived substantially  upon  the experimental demonstration of the correlation between an applied temperature
 gradient and the observed spin dynamics, including a spin current along the temperature gradient
 in an open-circuit magnetic sample, the so-called  spin Seebeck effect (SSE) \cite{UcTa08}.
  Meanwhile an impressive body work has accumulated on thermally induced spin- and spin-dependent currents
  \cite{UcTa08,HaBa09,BoSa11,JaYa10,UcXi10,XiBa10,UcNo10,HiNo11,ToMa12,BoGo13,JiBe11} (for a dedicated discussion we refer to the topical review \cite{AdUc13}).
The SSE was observed not only in metallic ferromagnets (FMs) like Co$_{2}$MnSi or semiconducting FMs, e.g. GaMnAs, (Ref. \cite{JaYa10}), but also in magnetic insulators LaY$_{2}$Fe$_{5}$O$_{12}$ (Ref. \cite{UcXi10})  and (Mn, Ze)Fe$_{2}$O$_{4}$ (Ref. \cite{UcNo10}).
The Seebeck effect is usually quantified by the Seebeck coefficient $S$ which is defined, in  a linear response manner,
as the ratio of the generated electric voltage $\Delta V$ to the temperature difference $\Delta T$: $S=-\frac{\Delta V}{\Delta T}$.
The magnitude of the Seebeck coefficient $S$ depends on the scattering rate and the density of electron states at the Fermi level, and thus it is material dependent variable. In the case of SSE, the spin voltage is formally determined by $\mu_{\uparrow}-\mu_{\downarrow}$, where $\mu_{\downarrow(\uparrow)}$ are the electrochemical potentials for spin-up and spin-down electrons, respectively. The density of states and the scattering rate for spin-up and spin-down electrons are commonly different, which results in various Seebeck constants for the two spin channels. In a metallic magnet subjected to a temperature gradient, one may think of the electrons in different spin channels to generate different driving forces, leading to a spin voltage that induces a nonzero spin current.
When a magnetic insulator is in contact with a normal metal (NM) and the system is subjected to a thermal gradient, the total spin current flowing through the interface is a sum of two oppositely directed currents. The current emitted from the FM into the NM, is commonly identified as a spin pumping current $I_{\mathrm{sp}}$ and originates from the thermally activated magnetization dynamics in the FM, while the other current $I_{\mathrm{fl}}$ is associated with the thermal fluctuations in the NM and is known as spin torque \cite{TsBr02}. The competition between the spin pump and the spin torque currents defines the direction of the total spin current which is proportional to the thermal gradient applied to the system. The theory of the magnon-driven SSE \cite{UcXi10} presupposes that the magnon temperature follows the phonon temperature profile and in a linear response approximation provides a good agreement with experiments.

In a recent study \cite{ChTo13} the theory of the magnon-driven SSE was extended beyond the linear response approximation. In particular, it was shown that the nonlinearity leads to a saturation of the total spin current and nonlinear effects become dominant when the following inequality holds $H_{0}/T_{\mathrm{F}}^{\mathrm{m}}<k_{\mathrm{B}}/(M_{\mathrm{s}}V)$, where $H_{0}$ is the constant magnetic field applied to the system, $T_{\mathrm{F}}^{\mathrm{m}}$ is the magnon temperature, $M_{\mathrm{s}}$ is the saturation magnetization and $V$ is the volume of the sample. The macrospin formulation of the stochastic Landau-Lifshitz-Gilbert (LLG) equation and the Fokker-Planck approach utilized in Ref. \cite{ChTo13} is inappropriate for non-uniformly magnetized samples with characteristic lengths exceeding several $10$ nm. Beyond the macrospin formulation the SSE effect for nonuniformly magnetized samples can be described by introducing a local magnetization vector \cite{BeSe09} $\vec{m}(\vec{r},t)$. In this case, however, the corresponding Fokker-Planck equation turns into an integro-differential equation and can only be solved after a linearization \cite{BoTr10}. Recently \cite{HoSa13}, the longitudinal SSE was studied in a NM-FM-NM sandwich structure in the case of a nonuniform magnetization profile. The linear regime, however, can not totaly embrace nontrivial and affluent physics of the SSE.

In the present study we inspect the SSE for a nonuniformly magnetized finite-size FM-NM interface subjected to an arbitrary temperature gradient.
Our purpose is to go beyond linear response regime which is relevant for the  nonlinear magnetization dynamics. It is shown that in analogy with the macrospin case \cite{ChTo13} the spin current in the nonlinear regime depends not only on the temperature gradient, but on the absolute values of the magnon temperature as well. In finite-size non-uniformly magnetized samples, however,
the site-dependent temperature profile may lead to  new physical important phenomena. For instance, we show that the key issue for the spin current flowing through a nonuniformly magnetized magnetic insulator is the local exchange spin torque and the local site-dependent magnon temperature profile, resulting in a generic spatial distribution
of the steady state spin current in a finite  chain subject to a uniform temperature gradient. The maximal spin current is predicted to be located at the middle of the chain.


\section{Theoretical framework}

For the description of the transversal magnetization dynamics we consider propagation of the normalized magnetization direction $\vec{m}(\vec{r},t)$ as
governed by the  Laundau-Lifshitz-Gilbert (LLG) equation \cite{LaLi35,Gilb55}
\begin{equation}
\label{LLG0}
\begin{split}
 \frac{\partial \vec{m}}{\partial t} = &-\gamma \left[\vec{m}\times\vec{H}^{\mathrm{eff}}\right]+\alpha \left[\vec{m}\times \frac{\partial \vec{m}}{\partial t}\right]\\ &-\gamma \left[\vec{m}\times \vec{h}(\vec{r},t)\right],
  \end{split}
\end{equation}
where the deterministic effective field  $\vec{H}^{\mathrm{eff}}=-\frac{1}{M_{\mathrm{S}}}\frac{\delta F}{\delta \vec{m}}$ derives
from the free energy density $F$ and is augmented by a Gaussian white-noise random field $h(\vec{r},t)$ with a space-dependent local intensity and  autocorrelation function. $\alpha$ is the Gilbert damping, $\gamma=1.76 \cdot 10^{11}$~[1/(Ts)] is the gyromagnetic ratio and $M_{\mathrm{S}}$ is the saturation magnetization.  $F$ reads
\begin{equation}
\label{eff}
\begin{split}
F=\frac{1}{V}\int \bigg[\frac{A}{2}|\vec{\nabla m}|^{2}+E_{a}(\vec{m})-\mu_0 M_{\mathrm{S}}\vec{H}_0\cdot\vec{m}\bigg]dV,
 \end{split}
\end{equation}
where $\vec{H}_0$ is the external constant magnetic field, $E_{\mathrm{a}}(\vec{m})$ is the anisotropy  energy density and $A$ is the exchange stiffness. $V$ is the system volume. We  employ a discretized version of the integro-differential equation (\ref{LLG0}) by defining
 $N$ cells with a characteristic length $a=\sqrt{2A/\mu_{0}M_{s}^{2}}$ of the exchange interaction between the magnetic moments. $a^{3}=\Omega_0$ is
  the  volume of the respective   cell.
  Assuming negligible variations of $\vec{m}(\vec{r},t)$ over a small $a$, one introduces
  a magnetization vector $\vec{M}_{n}$ averaged over the $n$th cell $\vec{M}_{n}=\frac{M_{\mathrm{S}}}{V}\int_{\Omega_0}\vec{m}(\vec{r},t)dV$
  and the total energy density becomes
\begin{equation}
\label{energydensity}
\begin{split}
  \displaystyle \varepsilon &=-\vec{H}_{\mathrm 0}\cdot \sum_n \vec{M}_n+\frac{K_1}{M^2_{\mathrm{S}}} \sum_n \left(M^2_{\mathrm{S}}-(M_{n}^{\mathrm z})^2\right) \\ & \quad-\frac{2A}{a^2 M^2_{\mathrm{S}}}\sum_n \vec{M}_n\cdot \vec{M}_{n+1}.
  \end{split}
\end{equation}
  $\vec{H}_{\mathrm 0}$ is the external magnetic field and $K_1$ is the uniaxial anisotropy density (with the easy axis: $\vec{e}_{\mathrm{z}}$).
   The effective magnetic field acting on the $n$-th magnetic moment reads
\begin{equation}
\label{effectivefieldT0}
\begin{split}
    \displaystyle \vec{H}_n^{\mathrm{eff}} &=-\frac{\partial \varepsilon}{\partial \vec{M}_n}=\vec{H}_{\mathrm 0}+\frac{2K_1}{M_{\mathrm{S}}^2}M_n^{\mathrm{z}}\vec{e}_{\mathrm{z}}
    \\ & \quad+\frac{2A}{a^2 M_{\mathrm{S}}^2}\left(\vec{M}_{n+1}+\vec{M}_{n-1}\right).
    \end{split}
\end{equation}
 Thermal activation is introduced by adding to the total effective field a stochastic fluctuating magnetic field  $ \vec{h}_n(t)$ so that
\begin{equation}
\label{effectivefield}
  \displaystyle \vec{H}^{\mathrm{eff}}_{n}(t)=\vec{H}_0+\vec{H}^{\mathrm{anis}}_n+\vec{H}^{\mathrm{exch}}_n+\vec{h}_n(t).
\end{equation}
Here $\vec{H}^{\mathrm{anis}}_{n}$ is the magnetic anisotropy field, $ \vec{H}^{\mathrm{exch}}_n $ is the exchange field. The random field $ \vec{h}_n(t)$ has a thermal origin and simulates the interaction of the magnetization with a thermal heat bath (cf. the review Ref. [\onlinecite{jap2012}] and references therein).
The site  dependence of $\vec{h}_{n}(t)$ reflects the existence of the local nonuniform temperature profile.  On the scale of the volume $\Omega_0$ the heat bath is considered uniform at a constant temperature.
The random field is characterized via the standard statistical properties of the correlation function
\begin{equation}
\label{eq_3}
     \begin{array}{l}
       \displaystyle{\left<h_{ik}(t)\right>=0,} \\
       \displaystyle{\left<h_{ik}(t)h_{jl}(t + \Delta t)\right> = \frac{2k_{B}T_i\alpha _i}{\gamma M_{\mathrm{S}}a^3} \delta_{ij} \delta_{kl} \delta(\Delta t).}
     \end{array}
\end{equation}
 $i$ and $j$ define the corresponding sites of the FM-chain and $k$, $l$ correspond to the cartesian components of the random magnetic field, $T_i$ and $\alpha_i$ are the site-dependent local temperature and the dimensionless Gilbert damping constant, respectively, $k_{\mathrm{B}}=1.38\cdot 10^{-23}$~[J/K] is the Boltzmann constant.

In what follows we employ for the numerical calculations the material parameters related to YIG, e.g. as tabulated  in Ref. \cite{XiBa10} (Table I).
 Explicitly the exchange stiffness is $A\approx 10$~[pJ/m], the saturation magnetization has a value of $4\pi M_{\mathrm{S}}\approx 10^6$~[A/m]. The anisotropy strength $K_1$ can be derived from the estimate for the frequency $\omega_0=\gamma 2 K_1/M_{\mathrm{S}}\approx 10\cdot 10^9$~[1/s] \cite{XiBa10}. The size of the FM cell is estimated from $a=\sqrt{2A/\mu_{0}M_{s}^{2}}$ yielding about 20 [nm]. For damping parameter we take the value $\alpha=0.01$, which exceeds the actual YIG value \cite{UcXi10,XiBa10}. This is done to optimize the numerical procedure in order to obtain reasonable calculation times. We note that although the quasi-equilibrium is assured when tracking the magnetization trajectories on the time scale longer than the relaxation time, the increased $\alpha$ quantitatively alters the strength in the correlation function (eq. (\ref{eq_3}))   and therefore indirectly has an impact on the values of the spin current.

\begin{center}
   \begin{figure}[!t]
    \centering
    \includegraphics[width=.45\textwidth]{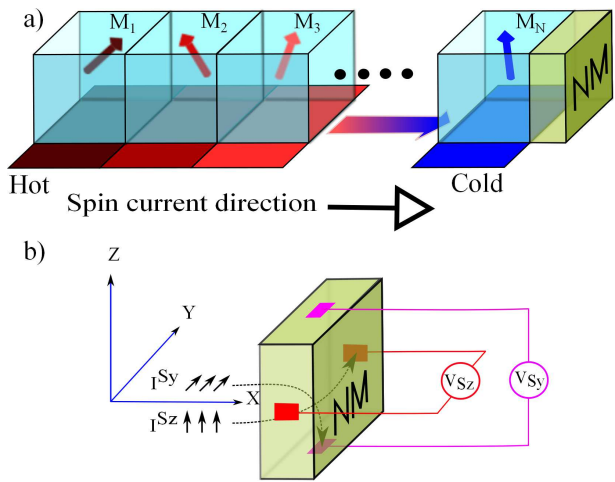}
        \caption{\label{fig_1} a) Schematics of the FM chain considered in the calculations. b) Suggested alignment for measurements.}
    \label{drawing}
    \end{figure}
\end{center}

We focus on a system representing a junction of a FM insulator and a NM which is schematically shown in FIG. \ref{drawing}. This illustration mimics the experimental setup for measuring the longitudinal SSE \cite{Kikkawa}, even though the analysis performed here does not include all the aspects of the experimental setting. The direction of the magnetic moments in the equilibrium is parallel to the FM-NM interface. Experimentally it was suggested to  pick up the longitudinal spin current by means of the inverse spin Hall effect  \cite{Kikkawa}. If it is so possible then, the electric field generated via the inverse spin Hall effect (ISHE) reads $\overrightarrow{\mathbf{E}}=D \overrightarrow{\mathbf{I_{s}}}\times\overrightarrow{\mathbf{\sigma}}$. Here  $\overrightarrow{\mathbf{E}}$ denotes electric field related to the inverse spin Hall effect, $\overrightarrow{\mathbf{I_{s}}}$ defines the spatial direction of the spin current, and $\overrightarrow{\mathbf{\sigma}}$ is spin polarization of the electrons in the NM,  and $D$ is the constant. We note, however, that our study is focused on the spin dynamics only and makes no statements on ISHE.

\section{Definition of the spin current}

For convenience we rewrite the Gilbert equation with the total energy density (\ref{energydensity})  in the form suggested in Ref. \cite{HoSa13}
\begin{equation}
\label{continuityequation}
\begin{split}
  \displaystyle \frac{\partial \vec{S}_n}{\partial t} &+\gamma \left[\vec{S}_n\times \left( \vec{H}^{\mathrm{eff}}_{n}(t)-\vec{H}^{\mathrm{ex}}\right)\right]+\frac{\alpha\gamma}{M_{\mathrm{S}}} \left[\vec{S}_n\times \frac{\partial \vec{S}_n}{\partial t}\right]
  \\&+\nabla\cdot\vec{J}^{\vec{s}}_n=0,
\end{split}
\end{equation}
where $\vec{S}_{n}=-\vec{M}_{n}/\gamma$ and the expression for the spin current density tensor reads
\begin{equation}
\label{nabla1}
  \displaystyle \nabla\cdot\vec{J}^{\vec{s}}_n=\gamma \left[\vec{S}_n\times\vec{H}^{\mathrm{ex}}_n\right].
\end{equation}
Here
\begin{equation}
\label{exchanget}
\vec{Q}_n=-\gamma [\vec{S}_n\times\vec{H}^{\mathrm{ex}}_n]
\end{equation} is the local exchange spin torque.

For the particular geometry (FIG.~\ref{drawing}) the only nonzero components of the spin current tensor are ${I}^{s_x}_{n} , {I}^{s_y}_{n} , {I}^{s_z}_{n}$. Taking into account eqs. (\ref{effectivefieldT0}) and (\ref{continuityequation}), we  consider a discrete version of the gradient operator and for the components of the spin current tensor $I_{n}^{s}=a^{2}J_{n}^{s}$ we deduce:
\begin{equation}
\label{longitudinalspincurrent}
 \displaystyle I^{\alpha}_{n}=I^{\alpha}_{0}-\frac{2Aa}{M_{S}^{2}}\sum_{m=1}^{n}M^{\mathrm{\beta}}_{\mathrm{m}}(M^{\mathrm{\gamma}}_{\mathrm{m-1}}+M^{\mathrm{\gamma}}_{\mathrm{m+1}})\varepsilon_{\alpha\beta\gamma},
\end{equation}
where $\varepsilon_{\alpha\beta\gamma}$ is the Levi-Civita antisymmetric tensor, Greek indexes define the current components and the Latin ones denote sites of the FM-chain. In what follows we will utilize eq. (\ref{longitudinalspincurrent}) for  quantifying the spin current in  the spin chain.
 We  consider different temperature gradients applied to the system taking into account the dependence of the magnon temperature on the phonon temperature profile \cite{UcXi10}.
  Since the temperature in the chain is not uniform, we expect a rich dynamics of different magnetic moments $\vec{M}_{n}$. In this case only nonuniform site-dependent spin current $I_{n}$ can fulfil the equation (\ref{continuityequation}). In order to prove this we will consider different configurations of magnetic fields for systems of different lengths. Modeling the interface effects between the FM insulator and the NM proceeds by invoking the concept of the enhanced Gilbert damping proposed in a recent study \cite{KaBr13}. The increased damping constant in the LLG equation of the last magnetic moment  describes losses of the spin current due to the interface effect. In order to evaluate the spin current flowing from the NM to the FM insulator we assume that the dynamics of the last spin in the insulator chain is influenced by the spin torque flowing from the NM to the magnetic insulator. The magnetic anisotropy is considered to have an easy axis \cite{UcXi10}.

\section{Numerical Results on isolated ferromagnetic insulator chain}

For the study of thermally activated magnetization dynamics we generate from $1000$ to $10000$ random trajectories for each magnetic moment of the FM-chain. All obtained observables are averaged over the statistical ensemble of stochastic trajectories. The number of realizations depends on the thermal gradient applied to the system. For long spin chains (up to $500$ magnetic moments) the calculations are computationally intensive  even for the optimized advanced numerical Heun-method \cite{KlPl91}, which converges in quadratic mean to the solution of the LLG equation when interpreted in the sense of Stratonovich \cite{Kamp07}. For the unit cell of the size $20$~[nm], the FM-chain of $500$ spins is equivalent to the magnetic insulator sample of the width around $10$ [$\mu$m]. We make sure  in our calculations that the magnetization dynamics is calculated on the large time scale exceeding the system's relaxation time which can be approximated via $\tau_{\mathrm{rel}}\approx M_{\mathrm{S}}/(\gamma 2 K_1 \alpha) \approx 10$~[ns]\cite{SuBe08}.

\subsection{Role of the local temperature and local spin exchange torque}

Prior to studying a realistic finite-size system we consider a toy model of  three coupled magnetic moments. Our aim is to better understand the role of local temperature and local exchange spin torque $Q_{n}$ (eq. (\ref{exchanget})) in the formation of the spin current $I_{n}$. Considering eqs. (\ref{nabla1}, \ref{exchanget}), we can utilize a recursive relation for the site-dependent spin current $I_{n}$ and the local exchange spin torque $I_{n}=I_{n-1}+\frac{a^3}{\gamma}Q_{n}$  for different temperatures of the site in the middle of the chain above $T_{2}>T_{\mathrm{av}}$ and below $T_{2}<T_{\mathrm{av}}$. The mean temperature in the system is $T_{\mathrm{av}}=\big(T_{1}+T_{2}+T_{3}\big)/3$.
The calculations are performed for different values of the site temperatures.
 We find that the exchange spin torques $Q_{n}$ related to magnetic moments $M_{n}$ with a temperature above the mean temperature $T_{n}>T_{\mathrm{av}}$ have a  positive  contribution to the spin current in contrast to the exchange spin torques $Q_{m}$ of the on-average-"cold" magnetic moments with $T_{m}<T_{\mathrm{av}}$.
 This  finding hints on the existence of a maximum spin current in a finite chain of magnetic moments and/or strong temperature gradient.
  This means that the site-dependent spin current $I_{n}$  increases if  $Q_{n}>0$ until the local site temperature drops below the mean temperature $T_{n}<T_{\mathrm{av}}$, in which
  case  the exchange spin torque becomes negative $Q_{n}<0$ and the spin current decreases.
  In order to prove that the negative contribution in the spin current of the on-average-cold magnetic moments is not an artefact of the three magnetic moments only,
   we studied long spin chains which  mimic non-uniformly magnetized magnetic insulators. In the thermodynamic limit for a large number of  magnetic moments $N\gg 1$ we expect to observe a formation of the equilibrium patterns in the spin current profile corresponding to the zero exchange spin torque $Q_{n}=0$ between nearest adjacent moments.

\subsection{Longitudinal spin current}

\begin{center}
   \begin{figure}[!t]
    \centering
    \includegraphics[width=.45\textwidth]{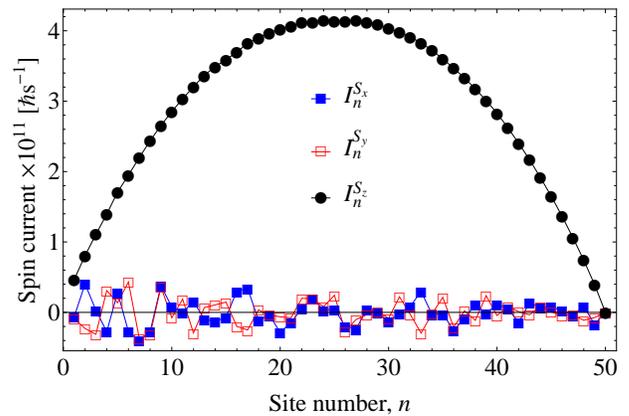}
        \caption{\label{fig2} {Different cartesian components of the statistically averaged longitudinal spin current as a function of the site number. Numerical parameters are $\Delta T=50$~[K], $\alpha =0.01$ and $H_0=0$~[T]}. The temperature gradient is defined $\Delta T=T_1-T_{50}$, where $T_1=50$~[K]. The only nonzero component of the spin current is $I_{n}^{S_{z}}$. Other two components $I_{n}^{S_{x}},~~I_{n}^{S_{y}}$ are zero because of the uniaxial magnetic anisotropy field which preserves $XOY$
        symmetry of the magnetization dynamics.}
    \label{fig1}
    \end{figure}
\end{center}

\begin{center}
   \begin{figure}[!t]
    \centering
    \includegraphics[width=.45\textwidth]{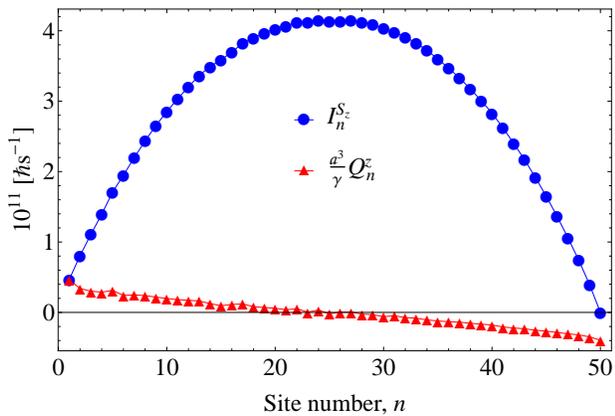}
        \caption{\label{fig_3} Z-component of the statistically averaged spin current $I_{n}^{S_{z}}$ (blue solid circles) and the distribution of the exchange spin torque $\frac{a^{3}}{\gamma}Q_{n}^{z}$ (red solid triangles), both site-dependent. Direct correlation between the behavior of the spin current and the exchange spin torque can be observed: the change of the sign of the exchange spin torque exactly matches the maximum of the spin current.}
    \label{exchange50}
    \end{figure}
\end{center}

\begin{center}
   \begin{figure}[!htb]
    \centering
    \includegraphics[width=.45\textwidth]{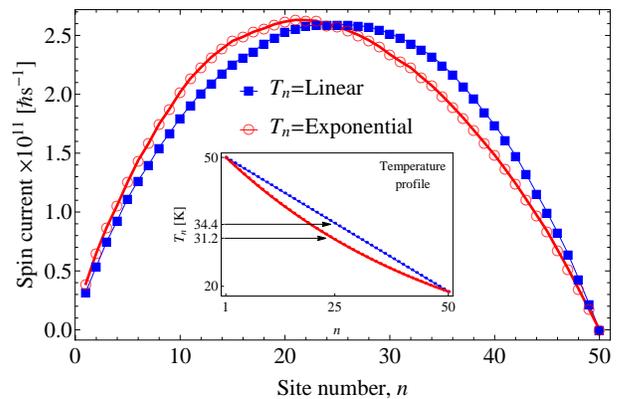}
        \caption{\label{fig_4} Z-component of the statistically averaged spin current for the linear $\Delta T=T_1-T_{50}$ and exponential $\Delta T(n)=50 \mathrm{[K]} \mathrm{e}^{-(n-1)/50}$ temperature gradients. The slight shift of the mean temperature to the left leads to a certain shifting of the maximum spin current to the left.}
    \label{exponential}
    \end{figure}
\end{center}

In FIG.~\ref{fig2} a dependence of distinct components of the spin current on the site is plotted. As inferred from the figure the current is not uniformly distributed along the chain. Evidently, the spin current has a maximum in the middle of the chain. The site-dependent spin current is an aftermath of the nonuniform magnon temperature profile applied to the system. This effect was not observed in the single macro spin approximation and is only relevant for the non-uniformly magnetized finite-size magnetic insulator sample. In addition one observes that the amplitude of the spin current increases with increasing the thermal gradient. This is predictably natural; less so  however, is the presence of a maximum of the spin current observed in the middle of the chain. We interpret this observation in terms of a collective cumulative averaged influence  of the surrounding  magnetic moments on particular magnetic moment. For a linear temperature gradient, as in  FIG.~\ref{fig1}, we have $\Delta T =\frac{T_{1}-T_{N}}{a N}$ which means that  half of the spins with $i<N/2$ possess  temperatures above the mean temperature of the chain $T_{1}/2$, while  the other half have  temperatures below the mean temperature. Further, the main contributors in the total spin current are the hot magnetic moments with temperatures above the mean temperature $T_{n}>T_{av}$ and with a positive exchange spin torque $Q_{n}>0$. While magnetic moments with a temperature below the mean temperature $T_{n}<T_{\mathrm{av}}$, $Q_{n}<0$ absorb the spin current and have a negative contribution in the total spin current. This non-equivalence of magnetic moments results in a maximum of the total spin current in the center of the chain. In what follows the magnetic moments with  temperatures higher than the mean temperature in the chain are  referred to as hot magnetic moments, while the magnetic moments with  temperatures lower than $T_{\mathrm{av}}$ we refer to as cold magnetic moments (i.e., our reference temperature is $T_{\mathrm{av}}$). The idea we are following is that the hot magnetic moments form the total spin current which is partly utilized for the activation of the cold magnetic moments.
FIG.~\ref{exchange50} illustrates the motivation of this statement. The maximum of the spin current (solid circles) is observed in the vicinity of the sites where the exchange spin torque term $Q_{n}$ changes its sign from  positive to  negative (solid triangles), highlighting the role of the hot and cold magnetic moments in  finite-size systems. To further affirm we consider two different temperature profiles - linear and exponential - with slightly shifted values of the mean temperature (FIG.~\ref{exponential}). The dependence of the maximum spin current on the mean temperature is a quite robust effect and a slight shifts of the mean temperature to the left lead to a certain shifting of the spin current's maximum. The effect of the nonuniform spin current passing through the finite-size magnetic insulator might be tested  experimentally using the SSE setup in which the spin current's direction is parallel to the temperature gradient. One may employ the inverse spin Hall effect using FM insulator covered by a stripe of paramagnetic metal, e.g. $Pt$ at different sites (cf Ref. \cite{Kikkawa}), albeit the chain must be small ($\lesssim 1\mu m$).

Furthermore, from FIG.~\ref{fig1} we infer that  the only nonzero component of the spin current is  $I_{n}^{z}$. Due to the uniaxial magnetic anisotropy
    all orientations of the magnetic moments in the $XOY$ plane are equivalent and $I_{n}^{\mathrm{x}}$,  $I_{n}^{\mathrm{y}}$ components of the spin current vanish.

\subsection{Role of boundary conditions}

\begin{center}
   \begin{figure}[!t]
    \centering
    \includegraphics[width=.45\textwidth]{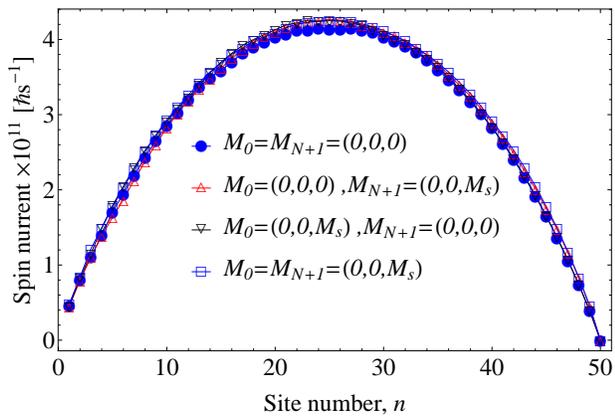}
        \caption{\label{fig_5} Effect of different boundary conditions on the averaged spin current. Numerical parameters are $\Delta T=50$~[K], $\alpha =0.01$ and $H_0=0$~[T]. The temperature gradient is defined $\Delta T=T_1-T_{50}$, where $T_1=50$~[K].
        In spite of different boundary conditions we observe the same maximal spin current for the site number corresponding to the
        mean temperature of the system.  Thus, the effect of the cold and hot magnetic moments is independent of the particular choice of boundary conditions.}
    \label{bound}
    \end{figure}
\end{center}

\begin{center}
   \begin{figure}[!t]
    \centering
    \includegraphics[width=.45\textwidth]{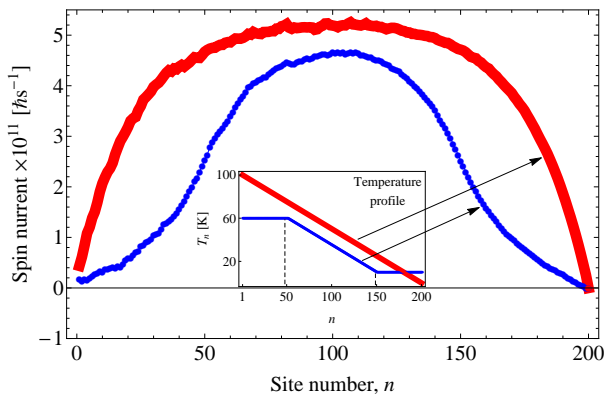}
        \caption{\label{fig_5b} Effect of boundary conditions in the case of different temperature profiles at the boundaries: linear temperature gradient (thick curve), constant temperature for $0<n<50$ and $150<n<200$ (thin curve). Even in the fragments of the chain with zero temperature gradient the spin current is not zero, which results from the formation of the spin current profile as a collective many body effect of the interacting magnetic moments. Therefore, the fragment of the chain with nonzero temperature gradient (sites $50<n<150$) has a significant influence on the formation of the spin current profiles in the left and right zero temperature gradient parts of the chain.}
    \label{bound_electrodes}
    \end{figure}
\end{center}

To elaborate on the origin of the observed maximum of the spin current we inspect the role of boundary conditions. In fact, in spite of employing different boundary conditions for the chain
we observe the same effect (FIG.~\ref{bound}), from which we can conclude
 that the effect of the cold and hot magnetic moments is inherent to the spin dynamics within the chain, which is independent from the particular choice of the boundary conditions. Furthermore,  we model the situation with the extended region at the ends of the FM chain (FIG.~\ref{fig_5b}), in which the  end temperatures are constant (i.e., one might imagine the heat reservoirs to have finite spatial extensions).  Modeling the ends of the FM-chain with zero temperature gradient by means of the LLG equations is certainly an approximation, which can be  improved by employing the Landau-Lifshitz-Bloch equations reported in Ref. \cite{AdUc13}. It captures, however, the main effects at relatively low temperatures:
  the flow  of the spin current for the decaying spin density away from the $T=\mathrm{const}$-$\Delta T$-interface and a non-zero integral spin current for the sites $0<n<50$ and $150<n<200$. As we see even in the fragments of the chain with a  zero temperature gradient the spin current is not zero.
  The reason is that the formation of the spin current profile is a collective many body effect of the interacting magnetic moments.
  Therefore, the  fragment of the chain with nonzero temperature gradient (sites $50<n<150$) has a significant influence on the formation of the spin current profiles in the left and right regions of the chain where  the temperature gradient vanishes.

\subsection{Temperature dependence of the longitudinal spin current}

\begin{center}
   \begin{figure}[!t]
    \centering
    \includegraphics[width=.45\textwidth]{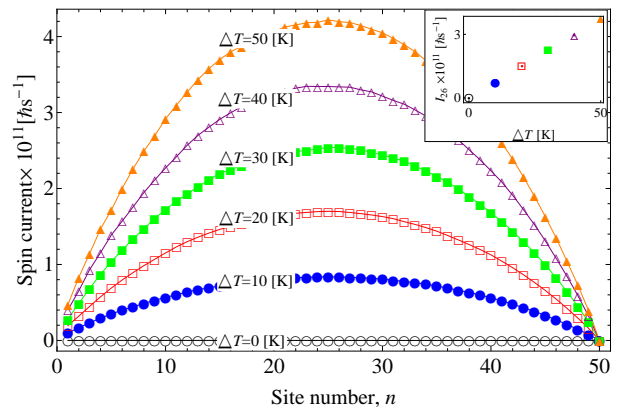}
        \caption{\label{fig_6} {Dependence of the averaged spin current on the strength of the temperature gradient. Numerical parameters are $\alpha =0.01$ and $H_0=0$~[T]}. The temperature gradient is defined as $\Delta T=T_1-T_{50}$, where $T_1=50$~[K]. The inset shows the averaged spin current for the $26$-th site. The maximum current increases with elevating the temperature gradient.}
    \label{differentDT}
    \end{figure}
\end{center}

In FIG.~\ref{differentDT} the dependence of the $z$-component of the averaged longitudinal spin current on the temperature gradient is shown. The dependence $I_n(\Delta T)$ (inset of FIG.~\ref{differentDT}) is linear and the amplitude of the spin current increases with the temperature gradient. This result is consistent with the experimental facts (Refs. \cite{JaYa10,UcXi10}) and our previous analytical estimations obtained via the single macrospin model \cite{ChTo13}.

\subsection{Finite-size  effects}

\begin{center}
   \begin{figure}[!t]
    \centering
    \includegraphics[width=.45\textwidth]{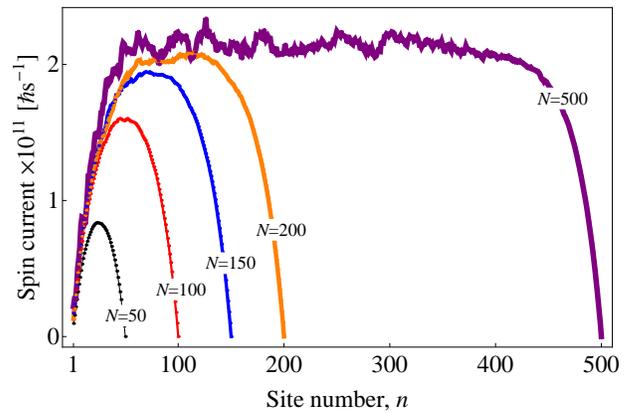}
        \caption{\label{fig_8} The dependence of the averaged spin current on the length of the FM-chain. Numerical parameters are $\alpha =0.01$ and $H_0=0$~[K]. The temperature gradient is linear and the maximum temperature is on the left-hand-side of the chain ($T_1=100$~[K]). In all cases the per-site temperature gradient is $\Delta T/N=0.2$~[K].}
    \label{finitesize}
    \end{figure}
\end{center}

\begin{center}
   \begin{figure}[!t]
    \centering
    \includegraphics[width=.45\textwidth]{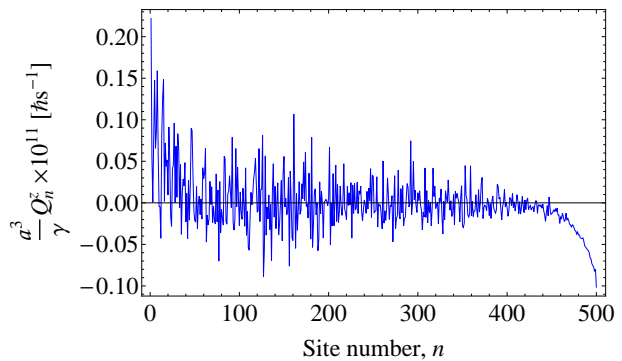}
        \caption{\label{fig_9} The dependence of the exchange spin torque on the site number. Numerical parameters are $\alpha =0.01$ and $H_0=0$~[K]. The temperature gradient is linear and the maximum temperature is on the left-hand-side of the chain ($T_1=100$~[K]). The per-site temperature gradient is $\Delta T/N=0.2$~[K]. The  exchange spin torque profile consists of three parts, the positive part corresponds to the high temperature domain and low temperature domain corresponds to the negative exchange spin torque. In the middle of the chain where the spin current is constant, the exchange spin torque fluctuates in the vicinity of the zero value.}
    \label{torque}
    \end{figure}
\end{center}

\begin{center}
   \begin{figure}[!t]
    \centering
    \includegraphics[width=.45\textwidth]{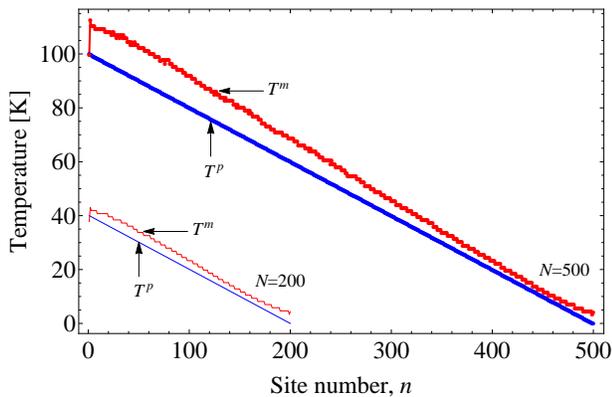}
        \caption{\label{fig_10} The magnon temperature profile (line) formed in the system. Numerical parameters are $\alpha =0.01$ and $H_0=0$~[K]. Blue line corresponds to the applied linear phonon temperature profile. The maximum temperature  on the left-hand-side of the chain is($T_1=100$~[K]). The per-site temperature gradient is $\Delta T/N=0.2$~[K]. The maximal deviation between the phonon and magnon temperatures is observed only at the left edge of chain. The difference between temperatures gradually decreases and becomes almost zero for the sites with $n>400$.}
    \label{magnonT}
    \end{figure}
\end{center}

Finite-size effects are considered relevant for the experimental observations (e.g. Ref. \cite{JaYa10}). In the thermodynamic limit  $N\gg1$ we expect the formation of  equilibrium patterns in the spin current profile corresponding to the zero exchange spin torque $Q_{n}=0$ between nearest adjacent moments. To address this issue, the spin current for chains of different lengths is shown in FIG.~\ref{finitesize}. As we see in the case of $N=500$ magnetic moments large pattern of the uniform spin current corresponding to the sites $50<n<450$ is observed.  In order to understand
such a behavior of the spin current for a large  system size, we plotted the dependence on the site number of the exchange spin torque $Q_{n}$ (FIG.~\ref{torque}). As we see, the exchange spin torque corresponding to the spin current plateau is characterized by large fluctuations around zero value, while nonzero positive (negative) values of the exchange spin torque $Q_{n}$ observed at the left (right) edges correspond to the nonmonotonic left and right wings  of the spin torque profile.
One may try to interpret the observed results in terms of the so called magnon relaxation length (MRL) $\lambda_{\mathrm{m}}\approx 2\sqrt{(D k_{\mathrm{B}}T/\hbar^2 )\tau_{mm}\tau_{mp}}$ (Refs. \cite{UcXi10,XiBa10}), where $D$ is the spin-wave stiffness constant and $\tau_{mm, mp}$ are the magnon-magnon, and the magnon-phonon relaxation times, respectively. The MRL is a characteristic length which results from the solution of the heat-rate equation for the coupled magnon-phonon system \cite{UcXi10}. The physical meaning of $\lambda_{\mathrm{m}}$ is an exponential drop of the space distribution of the local magnon temperature for the given external temperature gradient $\Delta T$. In other words, although the externally applied temperature bias is kept constant, the thermal distribution for magnons is not necessarily linear. In general, one may suggest a $\sinh(x)$-like spatial dependence \cite{UcXi10} and a temperature dependence  $\lambda_{\mathrm{m}}(T)$.
Estimates of the MRL for the material parameters related to YIG (suppl. mater. of Ref. \cite{UcXi10}) and $T_N=0.2$~[K] yield the following $\lambda_{\mathrm{m}}\approx 10$~[$\mu$m] \cite{estimate}. As seen from FIG.~\ref{finitesize} the length starting from which the saturation of the spin current comes into play as long as the FM-chain exceeds the length $20$~[nm]$\times 100\approx 2$~[$\mu$m]. However, we recall that MRL is a witness of the deviation between the magnon and phonon temperature profiles. Therefore,
 for interpreting the nonmonotonic parts of the spin current profile (FIG.~\ref{finitesize}) in terms of the MRL one has to prove the pronounced deviation between magnon and phonon temperatures at the boundaries. For further clarification we calculate the magnon temperature profile.
This can be done self-consistently via the  Langevin function $<M_{n}^{z}>=L\big(<M_{n}^{z}>H_{n}/k_{B}T_{n}^{m}\big)$. Here $H_{n}^{z}$ is the $z$ component of the local magnetic field which depends on the external magnetic field and the mean values of the adjacent magnetic moments $<M_{n-1}^{z}>,~~<M_{n+1}^{z}>$  ( see eq. (\ref{effectivefieldT0})). As inferred from the FIG.~\ref{magnonT}  the magnon temperature profile follows the phonon temperature profile. Prominent deviation between the phonon and the magnon temperatures is observed only at the beginning of the chain and gradually decreases and becomes small  on the MRL scale. Close to the end of the chain the temperature difference becomes almost zero. This means that left nonmonotonic parts of the spin current profile  FIG.~\ref{finitesize} can be interpreted in terms of none-equilibrium   processes. Comparing this result with the exchange spin torque profile (FIG.~\ref{torque}) we see that in this part of the spin chain the exchange spin torque is positive. This is the reason why the spin current $I_{n}$ is increasing with the site number $n$. The saturated plateau of the spin current shown in FIG.~\ref{finitesize} corresponds to the zero exchange spin torque $Q_{n}=0$ (cf. FIG.~\ref{torque}) and the decay of the spin Seebeck current $I_{n}$ at the right edge corresponds to the negative spin exchange torque $Q_{n}<0$. Thus, for the formation of the convex spin current profile the
key issue is not the difference between magnon and phonon temperatures, which as we see is pretty small, but the magnon temperature profile itself.
The existence of the hot(cold) magnetic moments with the local magnon temperature up (below) the mean magnon temperature generates the spin current.
This difference in the local magnon temperature of the  different magnetic moments drives the spin current in the chain. On the other hand any measurement of the spin current done in the vicinity of the right edge of the current profile will demonstrate a non-vanishing spin current in the absence of the deviation between the magnon and phonon temperature profiles. This may serve as an  explanation of the recent experiment \cite{Agrawal}, where a non-vanishing spin current was observed in the absence of the deviation between the magnon and the phonon temperature profiles. We note that zero values of the spin current shown in FIG.~\ref{finitesize} is the artefact of isolated magnetic insulator chain.  Real measurement of the spin currents usually involve FM-insulator/NM-interfaces. As will be shown below the interface effect described by an enhanced Gilbert damping and the spin torque lead to a nonzero spin current at the interfaces which is actually measured in the experiment.

\subsection{Role of the external magnetic filed ($H_0\neq0$)}

\begin{center}
   \begin{figure}[!t]
    \centering
    \includegraphics[width=.45\textwidth]{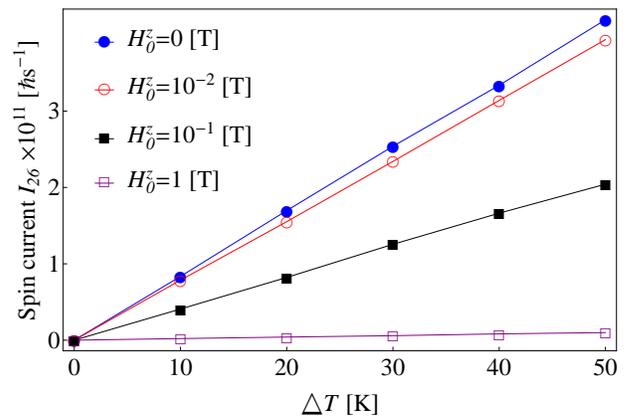}
        \caption{\label{fig_11} Effect of the external magnetic field applied parallel  to the easy axis on the averaged spin current. Numerical parameters are $\Delta T=50$~[K], $\alpha =0.01$ and $N=50$. The temperature gradient is linear and the maximum temperature is on the left-hand-side of the chain.}
    \label{BZ0dependency}
    \end{figure}
\end{center}

It follows from our calculations that the dependence of the longitudinal spin current on the magnetic field is not trivial. Once the external static magnetic field is applied perpendicularly to the FM-chain and along the easy axis at the same time,  we can suppress the spin current at elevated magnetic fields (FIG.~\ref{BZ0dependency}). The threshold magnetic field is - as expected - the strength of the anisotropy field, i.e. $2K_1/M_{\mathrm{S}}\sim0.056$~[T]. By applying  magnetic fields much higher than $0.056$~[T], the magnetic moments are fully aligned along the field direction and hence the X-, Y-components of the magnetization required to form the Z-component of the longitudinal averaged spin current vanish.

\begin{center}
   \begin{figure}[!t]
    \centering
    \includegraphics[width=.45\textwidth]{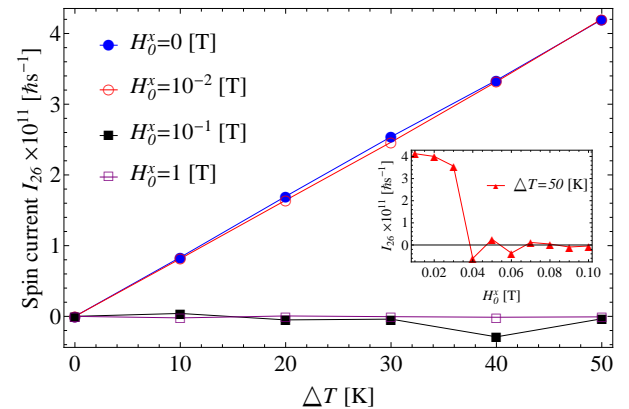}
        \caption{\label{fig_12} Effect of the external magnetic field applied perpendicularly to the easy axis on the averaged spin current. Numerical parameters are $\Delta T=50$~[K], $\alpha =0.01$ and $N=50$. The temperature gradient is linear and the maximum temperature is on the left-hand-side of the chain.}
    \label{BX0dependency}
    \end{figure}
\end{center}

In the case of the magnetic field being applied perpendicularly to the easy axis, the behavior becomes more rich (FIG.~\ref{BX0dependency}). In analogy with the situation observed in FIG.~\ref{BZ0dependency} there are no sizeable changes for the $I_n(\Delta T)$-dependence at low static fields. This is the regime where the anisotropy field is dominant. In contrast to the $H^{\mathrm{z}}_0$ applied field, the spin current does not linearly depend on the strength of the field (inset of FIG.~\ref{BZ0dependency}), which is explained by the presence of different competing contributions in the total energy density and not a simple correction of the Z-component of the anisotropy field illustrated in the previous figure. Surprisingly, the magnetic field oriented along the FM-chain can also suppress the appearance of the spin current's profile. Also in this case the strong magnetic field destroys the formation of the magnetization gradient resulting from the applied temperature bias.

\section{Interface effects}

The experimental setup to detect the spin current might involve a NM adjacent to the spin-current generating substance, e.g. a FM insulator. This NM converts the injected spin current from the FM to an electric current via ISHE \cite{UcTa08,UcXi10,SaUe06}. So it is of interest to see the effect of the adjacent NM on the generated spin current in the considered chain. Obviously, the main effects appear in the FM-NM interface. The interface effect can be divided into two parts which is described in the following subsections.

\subsection{Spin pumping and enhanced Gilbert damping}

In \emph{magnetic insulators}, charge dynamics is less relevant (in our model, anyway), and in some cases the dissipative losses associated with the magnetization dynamics are exceptionally low (e.g. in YIG\cite{KaHa10} $\alpha=6.7\times 10^{-5}$). When a magnetic insulator is brought  in contact with a \emph{normal metal}, magnetization dynamics results in spin pumping, which in turn causes angular momentum being pumped to the NM. Because of this nonlocal interaction, the magnetization losses become enhanced\cite{KaBr13}.

If we consider the normal metal as a \emph{perfect spin sink} which remains in equilibrium even though spins are pumped into it (which means there is a rapid spin relaxation and no back flow of spin currents to the magnetic insulator), the magnetization dynamics is described by the LLG equation with an additional torque originating from the FM-insulator/NM interfacial spin pumping\cite{KaBr13}
\begin{equation}
\label{LLG1}
  \displaystyle \frac{\partial \vec{M}}{\partial t}=-\gamma \left[\vec{M}\times \vec{H}^{\mathrm{eff}}\right]+\frac{\alpha}{M_{\mathrm{S}}} \left[ \vec{M}\times \frac{\partial \vec{M}}{\partial t} \right]+\vec{\tau}^{\mathrm{sp}},
\end{equation}
where
\begin{equation}
\label{LLG2}
  \vec{\tau}^{\mathrm{sp}}=\frac{\gamma \hbar}{4\pi M^2_{\mathrm{S}}}g_{\mathrm{eff}} \delta(x-L)\left[ \vec{M}\times \frac{\partial \vec{M}}{\partial t}\right],
\end{equation}
where $L$ is the position of the interface, $e$ is the electron charge and $g_{\mathrm{eff}}$ is the real part of the \textit{effective} spin-mixing conductance. In the YIG-Pt bilayer the maximum measured effective spin-mixing conductance is $g_{\mathrm{eff}}=4.8\times 10^{20}$~[m$^{-2}$] Ref. \cite{KaBr13}. In fact if the spin pumping torque should be completely described, one should add another torque containing the imaginary part of $g_{\mathrm{eff}}$ \cite{BrKe12}. However, we omit this imaginary part here because it has been found to be too small at FM-NM interfaces \cite{JiLi11}.

The aforementioned spin pumping torque concerns the cases that we characterized with $\vec{M}$. In our discrete model which includes a chain of $N$ ferromagnetic cells, we describe the above phenomena as follows
\begin{equation}
\label{LLG3}
  \displaystyle \frac{\partial \vec{M}_n}{\partial t}=-\gamma \left[ \vec{M}_n\times \vec{H}^{\mathrm{eff}}_{n}\right]+\frac{\alpha}{M_{\mathrm{S}}} \left[\vec{M}_n\times \frac{\partial \vec{M}_n}{\partial t}\right]+\vec{\tau}^{\mathrm{sp}}_n,
\end{equation}
where
\begin{equation}
\label{LLG4}
  \vec{\tau}^{\mathrm{sp}}_n=\frac{\gamma \hbar^2}{2a e^2M^2_{\mathrm{S}}}g_\bot \delta_{nN} \left[\vec{M}_n\times \frac{\partial \textbf{M}_n}{\partial t}\right],
\end{equation}
which means the spin pumping leads to an enhanced Gilbert damping in the last site
\begin{equation}
\label{enhanceddamping}
  \Delta\alpha=\frac{\gamma \hbar}{4\pi a M_s} g_{\mathrm{eff}}.
\end{equation}

As mentioned, the above enhanced Gilbert damping could solely describe the interfacial effects as long as we treat the adjacent normal metal as a perfect spin sink without any back flow of the spin current from the NM \cite{KaBr13,HoSa13}. The latter is driven by the accumulated spins in the normal metal. If we model the normal metal as a perfect spin sink for the spin current, spin accumulation does not build up. This approximation is valid when the spin-flip relaxation time is very small and so it prevents any spin-accumulation build-up. So the spins injected by pumping decay and/or leave the interface sufficiently fast and there won't be any backscattering into the ferromagnet \cite{TsBr02,TsBr05}.
We note by passing that in a recent study concerning this phenomena, it has been shown that spin pumping (and so enhanced Gilbert damping) depends on the transverse mode number and in-plane wave vector\cite{KaBr13}.

\subsection{Spin transfer torque}

It was independently proposed by Slonczewski \cite{Slon96} and Berger \cite{Berg96} that the damping torque in the LLG equation could have a negative sign as well, corresponding to a negative sign of $\alpha$. This means that the magnetization vector could move into a final position antiparallel to the effective field. In order to achieve this, energy has to be supplied to the FM system to make the angle between the magnetization and the effective field larger. This energy is thought to be provided by the injection of a spin current $\vec{I}^{\mathrm{incident}}$ to the FM \cite{ChToSu13,BrKe12,TsBr02}
\begin{equation}
\label{Slonczewski1}
  \vec{\tau}^{\mathrm{s}}=-\frac{\gamma}{M^2_{\mathrm{S}} V} \left[\vec{M}\times \left[\vec{M} \times \vec{I}^{\mathrm{injected}} \right]\right],
\end{equation}
which describes the dynamics of a monodomain ferromagnet of volume $V$ that is subject to the spin current $\vec{I}^{\mathrm{incident}}$ and modifies the right-hand side of the LLG equation as a source term. In general, a  torque-term additional  to the Slonczewski's torque (eq.~(\ref{Slonczewski1})) is also allowed \cite{BrKe12,StSi06}
\begin{equation}
\label{Slonczewski2}
  \vec{\tau}^{\mathrm{s\beta}}=-\frac{\gamma}{M_{\mathrm{S}} V}\beta \left[\vec{M} \times \vec{I}^{\mathrm{incident}}\right],
\end{equation}
where $\beta$ gives the relative strength with respect to the Slonczewski's torque (eq.~(\ref{Slonczewski1})).

For the case of a FM-chain, again we assume that the above spin-transfer torques act solely on the last FM cell.

\subsection{Numerical results for interface effects}
In order to simulate the enhanced Gilbert damping and the spin--transfer torque we assume that they act only on the chain end (motivated by their aforementioned origin). So the dynamics of our FM-chain is described by the following LLG \cite{LaLi35,Gilb55} equations
\begin{equation}
\label{LLG-N}
\begin{split}
  \displaystyle \frac{\partial \vec{M}_n}{\partial t} &=-\frac{\gamma}{1+\alpha^2} \left[\vec{M}_n\times \vec{H}^{\mathrm{eff}}_n\right]
  \\ & \quad-\frac{\gamma\alpha}{(1+\alpha^2)M_{\mathrm{S}}}\left[\vec{M}_n\times\left[\vec{M}_n\times \vec{H}^{\mathrm{eff}}_n\right]\right],
  \\ & \quad n=1,...,(N-1),
\end{split}
\end{equation}
and
\begin{equation}
\label{LLGN}
\begin{split}
  \displaystyle \frac{\partial \vec{M}_N}{\partial t} &=-\frac{\gamma}{1+\alpha_N^2} \left[\vec{M}_N\times \vec{H}^{\mathrm{eff}}_N\right]
  \\ & \quad-\frac{\gamma\alpha_N}{(1+\alpha_N^2)M_{\mathrm{S}}}\left[\vec{M}_N\times\left[\vec{M}_N\times \vec{H}^{\mathrm{eff}}_N\right]\right]
  \\ & \quad-\frac{\gamma}{M^2_{\mathrm{S}} a^3} \left[ \vec{M}_N\times \left[\vec{M}_N \times \vec{I}^{\mathrm{injected}} \right]\right]
  \\ & \quad-\frac{\gamma}{M_{\mathrm{S}} a^3}\beta \left[\vec{M}_N \times \vec{I}^{\mathrm{incident}}\right],
\end{split}
\end{equation}
where $\alpha_N=\alpha+\gamma \hbar g_{\mathrm{eff}}/(4\pi a M_s)$.

\begin{center}
   \begin{figure}[!h]
    \centering
    \includegraphics[width=.45\textwidth]{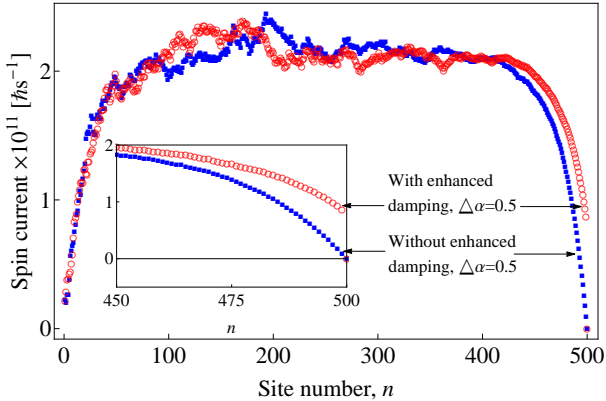}
        \caption{\label{fig_13} {Statistically averaged spin current in the chain of $N=500$-sites. Numerical parameters are $\Delta T=100$~[K], $\alpha =0.01$ and $H_0=0$~[T]. The temperature gradient is linear and the maximum temperature is on the left-hand-side of the chain ($T_1$). The blue curve shows the averaged spin current when no enhanced Gilbert damping and no spin--transfer torque is present. The red curve shows the averaged spin current when the enhanced Gilbert damping with $g_{\mathrm{eff}}=1.14\times 10^{22}$~[m$^{-2}$] is present. The inset shows the averaged spin current of the last fifty sites only.}}
    \label{500interface1-1}
    \end{figure}
\end{center}

\begin{center}
   \begin{figure}[!h]
    \centering
    \includegraphics[width=.45\textwidth]{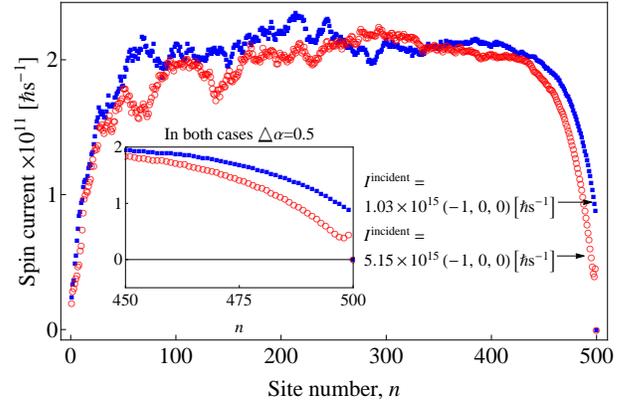}
        \caption{\label{fig_14} Statistically averaged spin current in the chain of $N=500$ when there are both the enhanced Gilbert damping and the spin--transfer torque. Numerical parameters are $\Delta T=100$~[K], $\alpha =0.01$, $H_0=0$~[T], $g_{\mathrm{eff}}=1.14\times 10^{22}$~[m$^{-2}$] and $\beta=0.01$. The temperature gradient is linear and the maximum temperature is on the left-hand-side of the chain ($T_1$). The blue curve has $\vec{I}^{\mathrm{incident}}=1\times 10^{15}(-1,0,0)$ $[\hbar \mathrm{s}^{-1}]$ and the red curve is with $\vec{I}^{\mathrm{incident}}=5\times 10^{15}(-1,0,0)$ $[\hbar \mathrm{s}^{-1}]$.}
    \label{500interface2-1}
    \end{figure}
\end{center}

Eq.~(\ref{LLG-N}) and (\ref{LLGN}) describe the magnetization dynamics in the presence of the interface effects and include both
spin pump and spin torque effects. Results in the absence of the spin torque are presented at the FIG.~\ref{500interface1-1}. The enhanced Gilbert damping captures losses of the spin current associated with the interface effect. A nonzero spin current
corresponding to the last $n=500$ spin quantifies the amount of the spin current pumped into the normal metal from the magnetic insulator.
However, the convex profile of the spin current is observed as well in the presence of the interface effects. The influence of the spin torque
on the spin current profile is shown in FIG.~\ref{500interface2-1}. We see from these results, the large spin torque reduces the total spin current following
through the FM-insulato/NM interfaces. The spin torque current is directed opposite to the spin pump current and therefore compensates it.

\section{Mechanisms of the formation of spin exchange torque and spin Seebeck current}

\begin{center}
   \begin{figure}[!h]
    \centering
    \includegraphics[width=.45\textwidth]{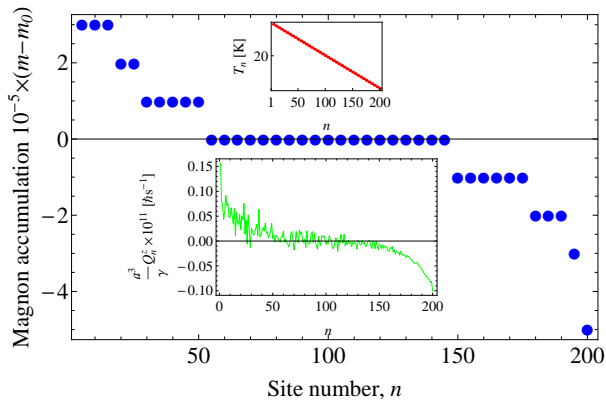}
        \caption{\label{fig_15} Site dependence of the exchange spin torque and
        the magnon accumulation effect. We observe a direct connection between magnon accumulation effect and the exchange spin torque.
 A positive magnon accumulation, i.e.  excess of the magnons, is observed in the high temperature part of the chain.
  While in the low temperature part  magnon accumulation is negative (lack of magnons compare to the equilibrium state).
   The exchange spin torque is positive for  positive magnon accumulation,
and  negative for  negative magnon accumulation. The spin Seebeck current is generated by excess magnons, transmitted through the equilibrium part of the chain and partially absorbed in the region with magnon drain.}
    \label{MagnonAccumulation}
    \end{figure}
\end{center}

In the previous sections we demonstrated the direct connection between the spin Seebeck current profile and the exchange spin torque. Here
we  consider the mechanisms of the formation of the exchange spin torque. For this purpose we investigate changes in the magnetization profile associated with the change of the magnon temperature  $<\Delta M_{n}^{z}>=< M_{n}^{z}>-< M_{0n}^{z}>$, where $< M_{n}^{z}>$
is the mean component of the magnetization moment for the case of the applied linear thermal gradient, while $< M_{0n}^{z}>$ corresponds to the mean magnetization component in the absence of thermal gradient $\Delta T=0$. Quantity $<\Delta M_{n}^{z}>$ defines  the magnon accumulation as the difference between the relative equilibrium  magnetization profile and excited one Ref. \cite{Nowak3} and is depicted in  FIG.~\ref{MagnonAccumulation}. We observe a direct connection between the magnon accumulation effect and the
exchange spin torque.
 A positive magnon accumulation, meaning an  excess of the magnons compared to the equilibrium state is observed in the high temperature part of the chain. While in the low temperature part the magnon accumulation is negative indicating
a lack of magnons compared to the equilibrium state. The exchange spin torque is positive in the case of the positive magnon accumulation and
is negative in the case of the negative magnon accumulation (the exchange spin torque vanishes in the equilibrium state).
From the physical point of view, the result is comprehensible: the spin Seebeck current is generated by the  magnon accumulation, transmitted through the equilibrium part of the chain and partially absorbed in the part of the chain with a negative magnon accumulation.

\section{Conclusions}

Based on the solution of the stochastic Landau-Lifshitz-Gilbert equation discretized for a ferromagnetic chain in the presence of a temperature gradient formed along the chain, we  studied the longitudinal spin Seebeck effect with a focus on the  space-dependent effects. In particular, we calculated a longitudinal averaged spin current as a function of different temperature gradients, temperature gradient strengths, distinct chain lengths and differently oriented external static magnetic fields. Our particular interest was to explain the mechanisms of the formation of the spin Seebeck current beyond the linear response regime. The  merit was in pointing out a microscopic mechanism for the emergence of the spin Seebeck current in a finite-size system. We have shown that, within our model, the microscopic mechanism  of the spin Seebeck current is the magnon accumulation effect quantified in terms of the exchange spin torque. We proved that the magnon accumulation effect drives the spin Seebeck current even in the absence of significant deviation between magnon and phonon temperature profiles. Our theoretical findings are in line with  recently observed experimental results \cite{Agrawal} where non-vanishing spin Seebeck current was observed in the absence of a temperature difference between phonon and magnon baths. 

Concerning the influence of the external constant magnetic fields on the spin Seebeck current we found that their role is nontrivial: An external static magnetic field applied perpendicularly to the FM-chain and along the easy axis may  suppress  the spin current at elevated magnetic fields (FIG.~\ref{BZ0dependency}). The threshold magnetic field   has a strength of the anisotropy field, i.e. $2K_1/M_{\mathrm{S}}\sim0.056$~[T]. In the case of the magnetic field applied perpendicularly to the easy axis, we observe a more complex behavior  (FIG.~\ref{BX0dependency}). In analogy with the situation seen in FIG.~\ref{BZ0dependency} there are no sizeable changes for the $I_n(\Delta T)$-dependence at low static fields. This is the regime where the anisotropy field is dominant. In contrast to the $H^{\mathrm{z}}_0$ applied field, it does not linearly depend on the strength of the field (inset of FIG.~\ref{BZ0dependency}), which is explained by the presence of different competing contributions in the total energy and not a simple correction of the Z-component of the anisotropy field. Notably, the magnetic field oriented along the FM-chain can also suppress the emergence  of the spin current's profile. Also in this case a strong magnetic field destroys the formation of the magnetization gradient resulting from the applied temperature bias.

In addition, we modeled an interface formed by a nonuniformly magnetized finite size ferromagnetic insulator  and a normal metal (e.g., YIG-Platinum junction) to inspect the effects of the enhanced Gilbert damping on the formation of  space-dependent spin current within the chain.

\section{Acknowledgements}
The financial support by the Deutsche Forschungsgemeinschaft (DFG) is gratefully acknowledged.


\begin{thebibliography}{99}
\bibitem{estimate}
We note that one should properly rescale the magnon-magnon and the magnon-phonon relaxation times given in Ref. \cite{UcXi10}, since we use in the numerical calculations the optimized values for damping. It is shown in Ref. \cite{XiBa10}, that the relaxation times are inversely proportional to the magnetization damping.
\end{thebibliography}
\end{document}